\documentclass[aps,prl,twocolumn,groupedaddress,showpacs]{revtex4}
\usepackage{graphicx}
\usepackage{dcolumn}
\usepackage{epstopdf}
\usepackage{bm}
\usepackage{amsfonts}
\usepackage{xcolor}

\newcommand{\ie}{i.\,e.\ }

\begin{document}

\title{Fitting in a complex $\chi^2$ landscape using an optimized hypersurface sampling}
\author{L.\,C.\ Pardo$^1$, M.\ Rovira-Esteva$^1$, S.\ Busch$^2$, J.-F.\ Moulin$^3$, J.\,Ll.\ Tamarit$^1$}

\affiliation{$^1$Grup de Caracteritzaci\'{o} de Materials,
Departament de F\'{\i}sica i Enginyieria Nuclear, ETSEIB,
Universitat Polit\`ecnica de Catalunya, Diagonal 647, 08028
Barcelona, Catalonia, Spain}
\affiliation{$^2$Physik Department E13 and
Forschungs-Neutronenquelle Heinz Maier-Leibnitz (FRM~II),
Technische Universit\"at M\"unchen,
Lichtenbergstr.\ 1, 85748 Garching,
Germany}
\affiliation{$^3$Helmholtz-Zentrum Geesthacht,
Institut f\"ur Werkstoffforschung,
Abteilung WPN, Instrument REFSANS,
Forschungs-Neutronenquelle Heinz Maier-Leibnitz (FRM~II),
Lichtenbergstr.\ 1, 85748 Garching,
Germany}

\begin{abstract}
Fitting a data set with a parametrized model can be seen geometrically as finding the global minimum of the $\chi^2$ hypersurface, depending on a set of parameters $\{P_i\}$. This is usually done using the Levenberg-Marquardt algorithm. The main drawback of this algorithm is that despite of its fast convergence, it can get stuck if the parameters are not initialized close to the final solution. We propose a modification of the Metropolis algorithm introducing a parameter step tuning that optimizes the sampling of parameter space. 
The ability of the parameter tuning algorithm together with simulated annealing to find the global $\chi^2$ hypersurface minimum, jumping across $\chi^2\{P_i\}$ barriers when necessary, is demonstrated with synthetic functions and with real data.
\end{abstract}

\pacs{02.50.Cw,02.50.Ng,02.60.Pn,02.50.Tt}

\maketitle

\section{Introduction}

Fitting a parametrized model to experimental results is the most usual way to obtain the physics hidden behind data. However, as nicely reported by Transtrum et al.~\cite{PRLfit}, this can be quite challenging and it usually takes ``weeks of human guidance to find a good starting point''. Geometrically, the problem of finding a best fit corresponds to finding the global minimum of the $\chi^2$ hypersurface. As this hypersurface is often full of fissures, local minima prohibit an efficient search. The human guidance consists usually of a set of tricks (depending on every particular problem) that allow to choose the starting point in this landscape such that the first minimum found is indeed the global minimum.

This problem is usually due to the mechanism that is behind classical fit algorithms such as Levenberg-Marquardt (LM)~\cite{numrecipes}: a set of parameters $\{P_i\}$ is optimized by varying the parameters and accepting the modified parameter set as a starting point for the next iteration only if this new set reduces the value of a cost or merit function such as $\chi^2$. From a geometrical point of view, those algorithms allow only downhill movements in the $\chi^2\{P_i\}$ hypersurface. Therefore they can get stuck in local minima or get lost in flat regions of the $\chi^2$ landscape~\cite{PRLfit}. This means that they are only able to find an optimal solution if they are initialized around the absolute minimum of the $\chi^2$ hypersurface. 

The challenge of finding the global minimum can be alternatively tackled by Bayesian methods~\cite{Bayes,Sivia_book} as demonstrated in different fields such as astronomy or biology~\cite{bayesapp}, solid state physics~\cite{bayesappCM}, quasielastic neutron scattering data analysis~\cite{Sivia_QENS}, and Reverse Monte Carlo methods~\cite{RMC}. We follow a Bayesian approach to the fit problem in this contribution. This method is based on another mechanism to wander around in parameter space: instead of allowing only downhill movements, parameter changes that increase $\chi^2$ can also be accepted if the change in $\chi^2$ is compatible with the data errors.

To do that, a Markov Chain Monte Carlo (MCMC) method is used, where the Markov Chains are generated by the Metropolis algorithm~\cite{hastings}. However, while in the case of the LM algorithm the initialization of parameters is critical to the convergence of the algorithm, it is here the tuning of the maximum parameter change allowed at each step (called parameter jumps hereafter) that will decide the success of the algorithm to find the global $\chi^2\{P_i\}$ minimum in an efficient way.

If the parameter jumps are chosen too small, the algorithm will always accept any parameter change, getting lost in irrelevant details of the $\chi^2\{P_i\}$ landscape. If chosen too large, the parameters will hardly be accepted and the algorithm will get stuck every now and then. Moreover, in the case of models defined by more than one parameter, when parameter jumps are not properly chosen, the parameter space can be over-explored in the direction of those parameters with too small jump lengths, in other words, the model would be insensitive to the proposed change of these parameters. On the other hand, some other parameters can be associated to a jump so big that changes are hardly ever accepted.

Different schemes have been proposed in order to change parameter jumps to explore the target distribution efficiently using Markov Chains under the generic name of adaptive MCMC \cite{Andrieu2008}. Using the framework of the Stochastic Approximation \cite{Benveniste1990} we present in this work an algorithm belonging to the group of ``Controlled Markov Chains'' \cite{Borkar1990,Andrieu2001} where the calculation of new parameter jumps takes the history of the Markov Chain and previous parameter jumps into account.

Two main approaches are known which take the Markov Chain history into account: Adaptive Metropolis (AM) algorithms\cite{Haario2001} (implemented for example in PyMC~\cite{PYMC}) and algorithms that use  rules following Robbins-Monro update \cite{Robbins1951,Gilks1998,Andrieu2001}. In the first case, parameter jumps are tuned using the covariance matrix at every step, so that once the adaptation is finished the algorithm should be wandering with a parameter jump close to the ``error'' of the parameter (defined as the variance of the posterior parameter PDF). In some cases, this kind of algorithm \cite{Andrieu2008} can get stuck if the acceptance ratio of a parameter is too high or too low. In this case the Markov Chain stops learning from the past history, thus the optimization is stopped with suboptimal parameter jumps. This problem is overcome by Robbins-Monro update rules that change parameter jumps so that they are accepted with an optimal ratio.

The main danger of optimized Metropolis algorithms is that adaptation might cause  the Markov Chain to not converge to the target distribution  anymore. In other words, the Markov Chain might lose its ergodicity. For example in the case of AM algorithms, the generated chain is not Markovian since it depends on the history of the chain. However, as demonstrated by Haario et al.~\cite{Haario2001}, the chain is able to reproduce the target distribution, i.e. is ergodic. In the second type of algorithms, the Robbins-Monro type, ergodicity properties must be assured by  updating  only at regeneration times \cite{Gilks1998}. In any case, as pointed out by Andrieu et al.~\cite{Andrieu2008} the convergence to the target distribution is assured if optimization vanishes. In other words, if parameter jumps oscillate around a fixed value the ergodic property of the Markov Chain is assured.

The presented algorithm is based on the stochastic approach of Robbins-Monro with an updating rule inspired by the one of Gilks et al.~\cite{Gilks1998}. Optimization of parameter jumps is therefore performed with two goals in mind:

\begin{itemize}
\item To calculate them in such a way that all parameters are accepted with the same ratio.
Adjusting parameter jumps so that all parameter changes will have the same acceptance ratio is important to explore the $\chi^2\{P_i\}$ landscape with the same efficiency in all parameter directions.

\item To adjust parameter jumps to a value tailored to the stage of the fit. This will turn out to be important when exploring the $\chi^2\{P_i\}$ hypersurface using the simulated annealing technique~\cite{Kirkpatrick1984}, since this allows the parameter jumps to be optimized to explore $\chi^2\{P_i\}$ (see subsection fitting in a complex $\chi^2$ landscape): 
at the beginning of the fit process the algorithm will set parameter jumps to a large value to explore large portions of the $\chi^2$ landscape, and at the final stages these parameter jumps will be set to small values by the same algorithm in order to find its absolute minimum.
\end{itemize}

Geometrically, we can interpret the algorithm as setting the parameter step sizes to a value related to the hypersurface landscape. First, it modifies the parameter jump to take into account the shape of
the hypersurface along a parameter direction. If $\chi^2\{P_{k}\}$  (the cut along a parameter $k$) is flat (the parameter direction is ``sloppy'' following Sethna's nomenclature~\cite{sloppy}), the parameter step size is set to a larger value, and parameters will move faster in this sloppy direction. On the contrary, in the directions where the $\chi^2\{P_{k}\}$ has a larger slope (the ``stiff'' direction following Sethna's nomenclature), parameter steps will be set to a smaller value so that they are accepted with the same as the previous ones. 
Second, it modifies the parameter jumps to take the shape of the global $\chi^2$ landscape into account when the simulated annealing is used. 
At the beginning of the fit parameter jumps will be set to a large value so that details of $\chi^2\{P_{k}\}$, i.e. local minima, will be smeared out, making it easier to find the global minimum. However, during the last steps of the fitting process, parameter steps will be set to a small value by the algorithm so that the system will be allowed to relax inside the minimum.

The present work gives a detailed description on how the algorithm works, and will be organized as follows: We first recall briefly on the Metropolis method applied to generate Markov Chains. In the next section, the proposed algorithm to optimize the parameter step size is introduced. Afterwards, we check its robustness to find optimized parameter jumps using a simple test function; and finally we test the ability of the regenerative algorithm combined with the simulated annealing technique to find the global minimum of $\chi^2$, even with poor initialization values, using a simple function with a complex $\chi^2\{P_i\}$ landscape. The algorithm presented in this work has been implemented in the program FABADA~\cite{fabada}.

\section{The fit method}
\subsection{Fitting with the Bayesian ansatz}

Fitting data using the Metropolis algorithm is based on an iterative process where successively proposed parameter sets are accepted according to the probability that these parameters describe the actual data, given all available evidence. Hence this method makes use of our knowledge of the error bars of the data.

We now briefly recall how this can be done using a Metropolis algorithm, to proceed in the next section with the algorithm to adjust parameter jumps.

We should first start with the probabilistic bases behind the $\chi^2$ definition. The probability $\mathbb{P}(H \mid D)$ that an hypothesis $H$ is correctly describing an experimental result $D$ is related to the likelihood $\mathbb{P}(D \mid H)$ that experimental data $D_k$ ($k=1,\ldots,n$) are correctly described by a model or hypothesis $H_k$ ($k=1,\ldots,n$); using Bayes theorem~\cite{Sivia_book,Bayes},
\begin{equation}
    \label{eqbayes}
    \mathbb{P}(H_k \mid D_k)=\frac{\mathbb{P}(D_k \mid H_k) \cdot \mathbb{P}(H_k)}{\mathbb{P}(D_k)}
\end{equation}
where $\mathbb{P}(H_k \mid D_k)$ is called the \emph{posterior}, the probability that the hypothesis is in fact describing the data. $\mathbb{P}(D_k \mid H_k)$ is the \emph{likelihood}, the probability that the description of the data by the hypothesis is good. $\mathbb{P}(H_k)$ is called the \emph{prior}, the probability density function (PDF) summarizing the knowledge we have about the hypothesis before looking at the data. $\mathbb{P}(D_k)$ is a normalization factor to assure that the integrated posterior probability is unity.

In the following we will assume no prior knowledge (maximum ignorance prior~\cite{Sivia_book}), in this special case Bayes theorem takes the simple form
\begin{equation}
\label{eqlikelidef}
    \mathbb{P}(H_k \mid D_k) \propto \mathbb{P}(D_k \mid H_k)\equiv L
\end{equation}
where $L$ is a short notation for likelihood.

Although this is by no means a prerequisite, we will assume in the following that the likelihood that every single data point $D_k$ described by the model or hypothesis $H_k$ follows a Gaussian distribution. The case of a Poisson distribution was discussed previously~\cite{fabada_paper}. For data with a Gaussian distributed uncertainty with width $\sigma$, the likelihood for each individual data point takes the form
\begin{equation}
\label{eqlikelig}
    \mathbb{P}(D_k \mid H_k)=\frac{1}{\sigma \sqrt{2\pi}} \exp\left[-\frac{1}{2}\left( \frac{H_k-D_k}{\sigma_k}\right)^2\right]
\end{equation}
and correspondingly, the likelihood that \emph{the whole} data set is described by this hypothesis is
\begin{eqnarray}
\label{eqlikeli}
    \mathbb{P}(D_k \mid H_k) &\propto& \prod_{k=1}^{n}\exp\left[-\frac{1}{2}\left( \frac{H_k-D_k}{\sigma_k}\right)^2\right] \nonumber \\
    &=& \exp\left[{-\frac{1}{2}\sum_{k=1}^{n} \left( \frac{H_k-D_k}{\sigma_k}\right)^2}\right] \nonumber \\
    &=& \exp\left(-\frac{\chi^2}{2}\right) \quad .
\end{eqnarray}

The Metropolis algorithm will in this special case consist on the proposition of successive sets of parameters $\{P_i\}$. A new set of parameters is generated changing one parameter at a time using the rule
\begin{equation}
\label{eqjump}
    P_i^\mathrm{new}=P_i^\mathrm{old}+ r \cdot \Delta P_i^\mathrm{max}
\end{equation}
where $\Delta P_i^\mathrm{max}$ is the maximum change allowed to the parameter or parameter jump and $r$ is a random number between -1.0 and 1.0. The new set of parameters will always be accepted if it lowers the value of $\chi^2$, or, if the opposite happens it will be accepted with a probability
\begin{equation}
    \label{eqaccept}
    \frac{\mathbb{P}(H\{P_i^\mathrm{l+1}\}\mid D_k)}{\mathbb{P}(H\{P_i^\mathrm{l}\}\mid D_k) } = \exp\left(-\frac{\chi^2_\mathrm{l+1}-\chi^2_\mathrm{l}}{2}\right)
\end{equation}
where $\chi^2_\mathrm{l+1}$ and $\chi^2_\mathrm{l}$ correspond to the  $\chi^2$ for the proposed new set of parameters and the old one, respectively. Otherwise, this new parameter value will be rejected and the fit function does not change during this step.

The Metropolis algorithm described here is very similar to the one used in statistical physics to find the possible molecular configurations (microstates) at a given temperature. In that case the algorithm minimizes the energy of the system while allowing changes in molecular positions that yield an increase of the energy if it is compatible with the temperature.

Inspired by the similarities between fitting data using a Bayesian approach and molecular modeling using Monte Carlo methods, a simulated annealing procedure proposed by Kirkpatrick  \cite{Kirkpatrick1984} might optionally be used (see for example \cite{Mortensen2005,Schulte1996}). Following the idea of that work, the $\chi^2$ landscape might be compared with an energy landscape used to describe glassy phenomena \cite{Debenedetti2001}. What we do is to start at high temperatures, i.e. in the liquid phase, where details of the energy landscape are not so important. By lowering the temperature fast enough the system might fall into a local minima, i.e. in the glassy phase. In that case the system is quenched as it is normally done by standard fitting methods. The presented algorithm aims to avoid being trapped in local minima using an "annealing schedule" as suggested by Kirkpatrick. This is done by artificially increasing the errors of the data to be fitted and letting the errors slowly relax until they reach their true values. Because this is very similar to what is performed in molecular modeling, the parameter favoring the uphill movements in equation \ref{eqacceptT} is usually called \emph{temperature}, yielding the acceptance rule
\begin{equation}
    \label{eqacceptT}
    \frac{\mathbb{P}(H(P_i^\mathrm{l+1})\mid D_k)}{\mathbb{P}(H(P_i^\mathrm{l})\mid D_k) }=\exp\left(-\frac{\chi^2_\mathrm{l+1}-\chi^2_\mathrm{l}}{2 \cdot T}\right) \quad .
\end{equation}

As it happens with Monte Carlo simulations, increasing the temperature will increase the acceptance of parameter sets that increase $\chi^2$, thus making the jump over $\chi^2$ barriers between minima easier. 

\subsection{Adjusting the parameter step size}

The objective of tuning the parameter step size is to choose a proper value for $\Delta P_i^\mathrm{max}$ in equation \ref{eqjump} to optimize the parameter space exploration.

Given the total number of algorithm steps $N$ and the number of steps that yield a change in $\chi^2$, \ie the number of successful attempts, $K$, the ratio $R$ of steps yielding a $\chi^2$ change is $R=K/N$. $R_\mathrm{desired}$ is defined as the ratio with which \emph{some parameter} should be accepted in a step. As we want every parameter to be changed with the same ratio, $R_{i,\mathrm{desired}}=R_\mathrm{desired}/m$ where $m$ is the number of parameters.

The algorithm is initialized with a first guess for the parameter step sizes. This first guess, as will be seen shortly, is not important due to the fast convergence of the algorithm to the optimized values. The calculation of a new $\Delta P_i^\mathrm{max}$, i.e. the regeneration of the Markov Chain, is done after $N$ steps, i.e. at regeneration times, through the equation
\begin{equation}
    \label{eqnewjump}
    \Delta P_{i}^\mathrm{max,new} = \Delta P_{i}^\mathrm{max,old} \cdot \frac{R_i}{R_{i,\mathrm{desired}}}
\end{equation}
where $R_i$ is the actual acceptance ratio of parameter $i$. Following the previous equation, if the calculated ratio $R_i / R_{i,\mathrm{desired}}$ is equal to one, \ie if all parameters are changing with the same predefined ratio, $\Delta P_{i}^\mathrm{max}$ will not be changed.

If during the fit process a change of parameter $P_i$ is too often accepted, the parameter space is being over explored with regard to parameter $i$. The algorithm will then make $\Delta P_{i}^\mathrm{max}$ larger in order to reduce its acceptance. The contrary happens if the acceptance is too low for a parameter: the algorithm makes $\Delta P_{i}^\mathrm{max}$ smaller to increase its acceptance ratio. This will set different step sizes for each parameter, making the exploration of all of them equally efficient.

\section{Demonstrations of fitting functions}

\subsection{Fitting in a well-behaved $\chi^2$ landscape}

\begin{figure}
\includegraphics[width=\columnwidth]{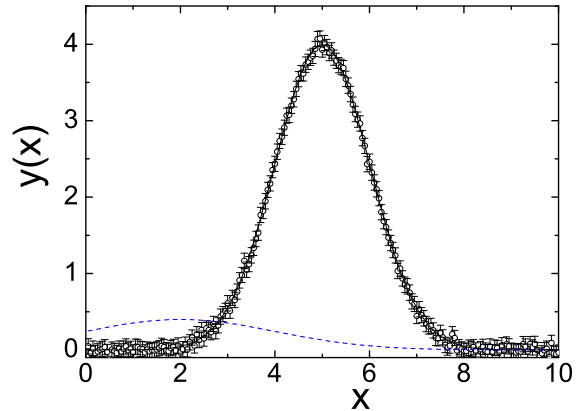}
\caption{Circles: Generated Gaussian function to test the algorithm with the parameters $\{A,W,C\}=\{10,1,5\}$. Dashed line: starting point for all performed tests ($\{A,W,C\}=\{2,2,2\}$). Solid line: best fit, \ie minimum $\chi^2$ fit, of the Gaussian function.}
\label{gausfit}
\end{figure}

The optimization of the parameter step size is shown using the Gaussian function

\begin{equation}
    \label{eqgauss}
    y(x) = \frac{A}{W \sqrt{2 \pi}} \exp  \left[ -\frac{(x-C)^2}{2W^2} \right]
\end{equation}
where $A$ is the amplitude, $W$ is the width and $C$ is the center of the Gaussian. A function has been generated with the parameter set $\{A,W,C\}=\{10,1,5\}$ and a normally distributed error with $\sigma=0.1$ was added. A series of tests with different initial values for parameter jumps and different desired acceptance ratios have been carried out (see below for details). The initial parameters for the fit were $\{A,W,C\}=\{2,2,2\}$. In all cases the algorithm was able to fit the data as can be seen in 
 \ref{gausfit}.

\begin{figure}
\includegraphics[width=\columnwidth]{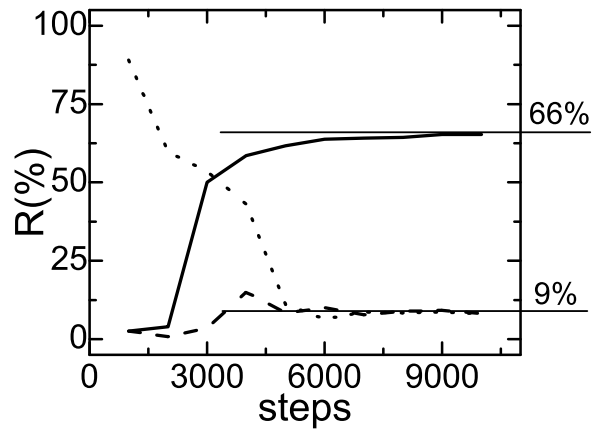}
\caption{Total acceptance ratio $R$ as a function of the number of steps when $R_\mathrm{desired}$ is set to 66\% and 9\% (solid and dashed or dotted lines). In the second case ($R_\mathrm{desired}=9\%$), dashed and dotted lines represent the values of $R$ as a function of algorithm step for two different parameter step size initializations ($\Delta P_{i}^\mathrm{max}=10$ and $\Delta P_{i}^\mathrm{max}=10^{-4}$ respectively) }
\label{totratio}
\end{figure}

The parameter step size was adjusted every 1000 steps. Three cases are shown in figure~\ref{totratio}: an initial $\Delta P_i^\mathrm{max}$ of 10 (a very large jump compared to the parameter values, nearly always resulting in a rejection of the new parameters) and an $R_\mathrm{desired}$ of 66\%, the same $\Delta P_i^\mathrm{max}$ with an $R_\mathrm{desired}$ of 9\% and finally a $\Delta P_i^\mathrm{max}$ of $10^{-4}$ (a very small jump compared to the parameter values, resulting in a slow exploration of the parameter space) and an $R_\mathrm{desired}$ of 9\%. It can be seen that the algorithm manages in all these extreme cases to adapt the jump size quickly and reliably in order to make $R$ equal to $R_\mathrm{desired}$.

\begin{figure}
\includegraphics[width=\columnwidth]{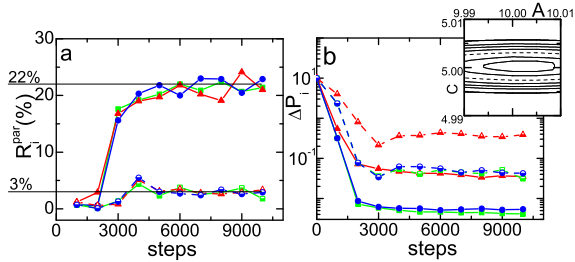}
\caption{(color online) a) Acceptance ratio $R_i$ for parameters $A$, $W$, $C$ involved in the fit of the Gaussian following equation \ref{eqgauss} ( red triangles, green squares and blue circles respectively) when $R_\mathrm{desired}$ is set to 66\% and 9\% (solid and dashed lines). b) Parameter step size as a function of the number of steps (line and symbols code as in figure a). The inset shows a cut through the $\chi^2$ hypersurface along $A$ and $C$ directions fixing W to the best fit value. }
\label{ratio}
\end{figure}

In figure \ref{ratio} we show the three individual acceptance ratios $R_i$ for the different parameters as a function of the fit steps for different initialization values of the parameter jumps $\Delta P_{i}$, for different values of $R_\mathrm{desired}$, and setting the number of steps to recalculate parameter jumps $N$ to 1000. When the total acceptance ratio is set to $R_\mathrm{desired}=66\%$ (solid line), the algorithm is able to change all parameter jumps (see figure \ref{ratio}(b)), making the acceptance ratio $R_i$ of every parameter equal to $R_\mathrm{desired}/m=22\%$ and thus the total acceptance ratio $R$  to 66\%. The same happens if the acceptance is set to 9\%: the algorithm finds the parameter step sizes (see dashed line in Fig. \ref{ratio}(b)) which yield a total acceptance ratio of 9\% within the first 5000 steps, no matter how the parameter step sizes were initialized. 

To explicitly show how this is linked with the geometrical features of the $\chi^2$ landscape, the inset of figure \ref{ratio}(b) shows a cut of the $\chi^2$ hypersurface along parameters $A$ and $C$, leaving parameter $W$ fixed to its best fit value $W_\mathrm{BF}$. As can readily be seen, the $\chi^2\{A,C,W=W_\mathrm{BF}\}$ hypersurface is sloppy in the direction of parameter $A$ and stiff in the direction of parameter $C$. The algorithm has thus correctly calculated a parameter step size which is larger for $A$ than for $C$, along whose direction the $\chi^2$ well is narrower. This fact makes the final parameter step sizes proportional to the errors of each parameter -- if the global minimum is not multimodal, is quadratic in all parameters, and those are not correlated. 

\begin{figure}
\includegraphics[width=\columnwidth]{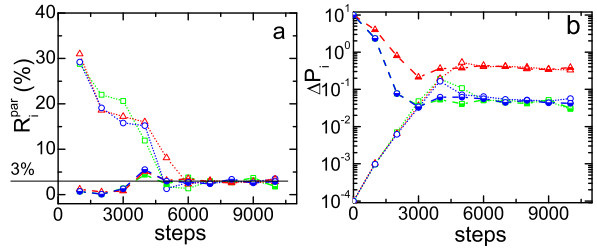}
\caption{(color online) a) Acceptance ratio $R_i$ for parameters $A$ (triangles), $W$ (squares), $C$ (circles) involved in the fit of the Gaussian following equation \ref{eqgauss} when initial parameter step sizes are set to  $\Delta P_i=10$ (dashed line) and $\Delta P_i=10^{-4}$ (dotted line). b) Parameter step size as a function of the number of steps (lines and symbols as in figure a).}
\label{initia}
\end{figure}

In order to show the robustness of the algorithm, we have also made disparate initial guesses for parameter step sizes $\Delta P_{i}^\mathrm{max}$ about three decades below the correct acceptance ratio, setting $R_\mathrm{desired}=9\%$. As displayed in figure \ref{ratio}, after about 5000 steps the acceptance ratio $R$ ($N$ is again 1000 steps) has already reached the desired value. It can be seen in figure \ref{initia}(a) that the acceptance ratio for each parameter reaches again the value $R_\mathrm{desired}/m=3\%$ and parameter step sizes are virtually equal to those obtained previously as shown in figure \ref{initia}(b).

\begin{figure}
\includegraphics[width=\columnwidth]{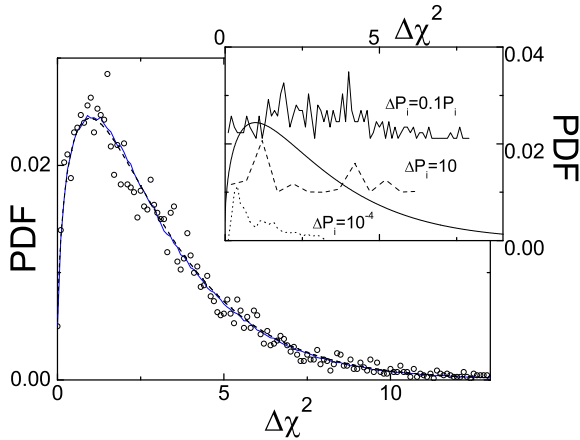}
\caption{The dashed line represents a chi-square distribution for three parameters, \ie $m=3$ (see text for details). Solid line is the obtained PDF associated to $\Delta\chi^2$
when calculated for $10^5$ steps. Circles represent the same distribution when calculated using only $10^4$ steps. The inset shows the $\chi^2$ PDFs when calculated with parameters allowed to change with $\Delta P_i = 10^{-4}$, $\Delta P_i = 10$, $\Delta P = 0.1P_i$. Successive PDFs are displaced on the ordinate axis for clarity of the figure.}
\label{acceptancechi}
\end{figure}

To stress the relevance of the aforementioned algorithm to explore the parameter space correctly, thus assuring its convergence, we have calculated the normalized $\Delta\chi^2$PDF in all tested cases. As can be seen in figure \ref{acceptancechi}, the $\Delta\chi^2$ PDF after $10^5$ steps matches the chi-square distribution

\begin{equation}
    \label{chisqdistri}
    \mathbb{P}(\Delta\chi^2) \propto \left(\Delta\chi^2\right)^{\left(\frac{m}{2}-1\right)}\exp\left(-\frac{\Delta\chi^2}{2}\right)
\end{equation}
with $m=3$ as expected~\cite{numrecipes}. In figure \ref{acceptancechi} we show the $\Delta\chi^2$ PDF obtained after $10^4$ steps for different cases: first setting $\Delta P_{i}^\mathrm{max}$ equal to the value calculated by the algorithm and second setting $\Delta P_{i}^\mathrm{max}$ equal to the initial guess and finally to a value, calculated a posteriori, which is proportional to the best fit parameters $\Delta P_{i}^\mathrm{max}=0.1P_i$ (inset of figure \ref{acceptancechi})

As can be seen in figure \ref{acceptancechi}, when $\Delta P_{i}^\mathrm{max}$ is set much higher than the optimal step sizes, the Metropolis algorithm scans the whole parameter space $\{P_i\}$, but jumping between disparate regions with very different values of $\chi^2$, therefore with a low acceptance rate of new parameter sets (dashed line in figure \ref{acceptancechi}). This causes a poor exploration of parameter space. In contrast, a small value over-explores only a restricted portion of $\{P_i\}$, falling very often in local minima of the parameter space (dotted line in the same figure). Also choosing parameter jumps proportional to the final parameters leads to a poor exploration of parameter space (solid line in the same figure). Finally, after the same number of steps, when using the optimized parameter step sizes obtained by the algorithm the $\chi^2$ PDF follows the theoretical expectation, meaning that the parameter space is correctly sampled. 

\subsection{Fitting in a complex $\chi^2$ landscape}

\begin{figure}
\includegraphics[width=\columnwidth]{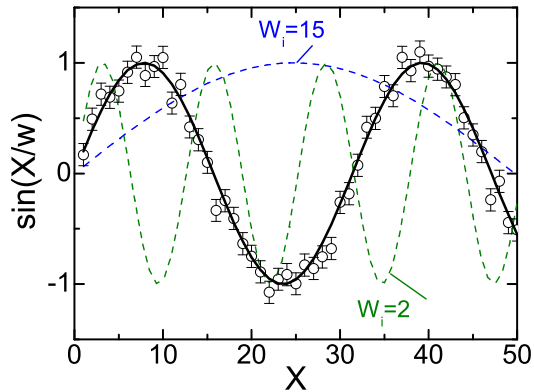}
\caption{(color online) Synthetic $\sin(x/5)$ function (circles) together with the best fit using parameter step sizes tuning together with simulated annealing (line). Dashed lines are the fits using the LM algorithm with starting parameters $W_i=2$ and $W_i=15$.}
\label{seno}
\end{figure}

\begin{figure}
\includegraphics[width=\columnwidth]{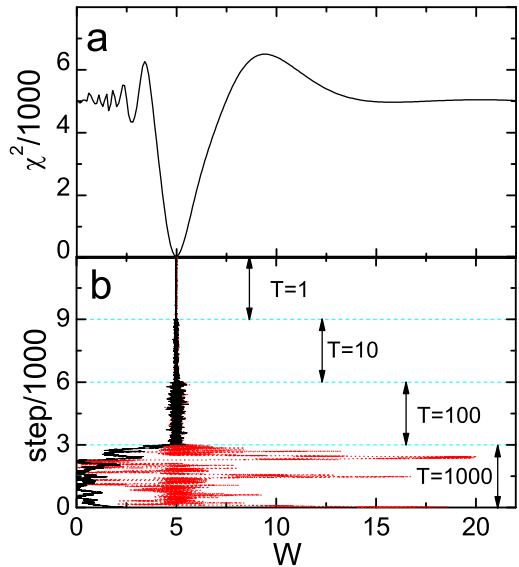}
\caption{(color online) (a) $\chi^2\{W\}$ landscape obtained for the function $\sin(x/W)$ with a normal error associated of $\sigma=0.1$ (see figure~\ref{seno}). (b) Algorithm steps for two different initializations , black solid line for $W_i=2$  and red dashed line for $W_i=15$, as a function of parameter $W$}
\label{sinuchi}
\end{figure}

As pointed out before, one of the main problems when dealing with data fitting using the LM algorithm is to find a proper set of initial parameters close enough to the global minimum of the $\chi^2\{P_i\}$ hypersurface. As an example we show in figure \ref{seno} the function $\sin(x/W)$ for $W=5$ affected by a normal distributed error with $\sigma=0.1$. In figure \ref{sinuchi}(a) we show the $\chi^2\{W\}$ landscape associated to the generated function. As it can be seen, the $\chi^2\{W\}$ landscape for this function has a great number of local minima and a global minimum at $W=5$. We have fitted the function using the LM algorithm and initializing the parameter at $W_i=2$ and $W_i=15$ (see figure \ref{seno}). As expected, both fits were not able to find the global minimum that fits the function. In fact only if the LM algorithm is initialized between $W=3.6$ and $W=9.0$ it is able to succeed in fitting the data.

We now test the ability of our algorithm to jump across $\chi^2$ barriers delimiting successive local minima to find the global one. For this task we have used the simulated annealing method, decreasing the temperature one decade every 3000 steps from $T=1000$ to $T=1$. The parameter jump calculation has been performed every $N=1000$ steps. While the initial temperature allows to explore wide regions of the parameter space, the last temperature will let the acceptance be determined only by the real errors of the data.

In figure \ref{sinuchi}(b) we show the parameter $W$ as a function of algorithm step for the two aforementioned initializations together with the $\chi^2$ landscape (a). Parameter step sizes were initialized after a first run of optimization of 2000 steps. As can be seen in this figure, after 3000 steps both runs have already reached the absolute $\chi^2$ minimum. Successive steps just relax the system to the final temperature $T=1$.

As it can be seen in figure \ref{sinuchi}, the way the minimum is reached depends on the parameter initialization. Parameter step sizes are larger for the run started with $W_i=15$ with a flat local minimum. The contrary happens with the run initialized at $W_i=2$, parameter step sizes are set small due to the narrow wells of the $\chi^2$ landscape in this region. However, both runs are able to avoid getting stuck in local minima, jumping over rather high $\chi^2$ barriers and successfully reaching the best fit. 

\section{Conclusion}

Classical fit schemes are known to fail when the parameters are not initialized close enough to the final solution. We have proposed in this work to use an Adaptive Markov Chain Monte Carlo Through Regeneration scheme, adapted from that of Gilks et al.~\cite{Gilks1998}, combined with a simulated annealing procedure to avoid this problem. 

The proposed algorithm tunes the parameter step size in order to assure that all of them are accepted in the same proportion. Geometrically the parameter step size is set large when a cut of $\chi^2\{P_i\}$ along this parameter is flat, \ie when the change of the $\chi^2\{P_i\}$ hypersurface along this parameter is sloppy. Similarly the parameter step size is set small if $\chi^2\{P_i\}$ wells are narrow.

Moreover, the step sizes can be modulated by a temperature added to the acceptance equation that makes jumps across $\chi^2$ barriers easier, \ie using a simulated annealing method~\cite{Kirkpatrick1984}. From a geometric point of view, a high temperature makes the $\chi^2\{P_i\}$ wells artificially broader, smearing out details of local minima. This is important at the first stages of a fit process. At final stages of the fitting, temperature is decreased, making parameter jumps smaller, and thus allowing the system to relax, once it is inside the global minimum.

By fitting simulated data including statistical errors we verified that our algorithm actually fulfills the  requirements of ergodicity (it converges to the target distribution), robustness (the ability to reach the $\chi^2$ minimum independent of the choice of starting parameters), ability to escape local minima and to explore efficiently the $\chi^2$ landscape, and guarantee that it will self tune to converge to the global minimum avoiding an infinite search with large steps.

More complex problems have already successfully been studied with this algorithm such as model selection using Quasielastic Neutron Scattering data~\cite{QENS}, non-functional fits in the case of dielectric spectroscopy~\cite{dielectric} or finding the molecular structure from diffraction data with a model defined by as many as 27 parameters~\cite{freon}. In the last case, the proper initialization of parameters to use a LM algorithm would have been a difficult task, made easy by the use of the presented algorithm.

\section{Acknowledgments}
This work was supported by the Spanish Ministry of Science and Technology
(FIS2008-00837) and by the Catalonia government (2009SGR-1251). We would also like to thank helpful comments and discussions on the manuscript made from K. Kretschmer,Anand Patil and Christopher Fonnesbeck and A. Font.

\end{document}